\documentclass[preprint]{emulateapj}

\usepackage{amsmath}

\slugcomment{To be Published in the Astrophysical Journal}
\shortauthors{Zaritsky, et al.}
\shorttitle{}

\begin{document}
\title{Globular Cluster Populations: Results Including S$^4$G Late-Type Galaxies}
  
\author{Dennis Zaritsky\altaffilmark{1}, Kelsey McCabe\altaffilmark{1}, Manuel Aravena\altaffilmark{2}, E. Athanassoula\altaffilmark{3}, Albert Bosma\altaffilmark{3}, S\'ebastien Comer\'on\altaffilmark{4,5}, Helene M. Courtois\altaffilmark{6,7}, Bruce G. Elmegreen\altaffilmark{8}, Debra M. Elmegreen\altaffilmark{9}, Santiago Erroz-Ferrer\altaffilmark{10}, Dimitri A. Gadotti\altaffilmark{11}, 
Joannah L. Hinz\altaffilmark{12}, Luis C. Ho\altaffilmark{13,14}, Benne Holwerda\altaffilmark{15}, Taehyun Kim\altaffilmark{16,17}, Johan H. Knapen\altaffilmark{18,19}, 
Jarkko Laine\altaffilmark{4}, 
Eija Laurikainen\altaffilmark{4,5}, 
Juan Carlos Mu\~noz-Mateos\altaffilmark{11,16}, Heikki Salo\altaffilmark{4},  and Kartik Sheth\altaffilmark{16,20}}

\altaffiltext{1}{Steward Observatory, University of Arizona, 933 North Cherry Avenue, Tucson, AZ 85721, USA; dennis.zaritsky@gmail.com}
\altaffiltext{2}{N\'ucleo de Astronom\'{\i}a, Facultad de Ingenier\'{\i}a, Universidad Diego Portales, Av. Ej\'ercito 441, Santiago, Chile}
\altaffiltext{3}{Aix Marseille Universit\'e, CNRS, LAM (Laboratoire d'Astrophysique de Marseille) UMR 7326, 13388, Marseille, France}
\altaffiltext{4}{Astronomy and Space Physics, P.O. Box 3000, FI 90014 University of Oulu, Finland}
\altaffiltext{5}{Finnish Centre of Astronomy with ESO (FINCA), University of Turku, V\"ais\"al\"antie 20, FI-21500, Piikki\"o, Finland}
\altaffiltext{6}{Universit\'e Lyon 1, CNRS/IN2P3, Institut de Physique Nucl\'eaire, Lyon, France}
\altaffiltext{7}{Institute for Astronomy, University of Hawaii, 2680 Woodlawn Drive, Honolulu, H{\small I} 96822, USA}
\altaffiltext{8}{IBM T. J. Watson Research Center, 1101 Kitchawan Road, Yorktown Heights, NY 10598, USA}
\altaffiltext{9}{Vassar College, Dept.  of Physics and Astronomy, Poughkeepsie, NY 12604}
\altaffiltext{10}{Department of Physics, Institute for Astronomy, ETH Zurich, CH-8093 Zurich, Switzerland}
\altaffiltext{11}{European Southern Observatory, Casilla 19001, Santiago 19, Chile}
\altaffiltext{12}{MMT Observatory, P.O. Box 210065, Tucson, AZ 85721, USA}
\altaffiltext{13}{Kavli Institute for Astronomy and Astrophysics, Peking University, Beijing 100871, China}
\altaffiltext{14}{Department of Astronomy, Peking University, Beijing 100871,China}
\altaffiltext{15}{University of Leiden, Leiden Observatory, Niels Bohrweg 4, NL-2333, Leiden, The Netherlands}
\altaffiltext{16}{National Radio Astronomy Observatory/ NAASC, 520 Edgemont Road, Charlottesville, VA 22903, USA }
\altaffiltext{17}{Korea Astronomy and Space Science Institute, Daejeon 305-348, Republic of Korea}
\altaffiltext{18}{Instituto de Astrof\'isica de Canarias, V\'{i}a L\'actea, S/N, 38205, La Laguna, Spain}
\altaffiltext{19}{Departamento de Astrof\'{\i}sica, Universidad de La Laguna, 38206, La Laguna,  Spain}

\begin{abstract} 
Using 3.6 and 4.5$\mu$m images of 73 late-type, edge-on galaxies from the S$^4$G survey, we compare the richness of the globular cluster populations of these galaxies to those of early type galaxies that we measured previously.  In general, the galaxies presented here fill in the distribution for galaxies with lower stellar mass, M$_*$, specifically $\log({\rm M}_*/{\rm M}_\odot) < 10$, overlap the results for early-type galaxies of similar masses, and, by doing so, strengthen the case for a dependence of the number of globular clusters per $10^9\  {\rm M}_\odot$ of galaxy stellar mass, T$_{\rm N}$, on M$_*$. For $8.5 < \log ({\rm M}_*/{\rm M}_\odot) < 10.5$ we find the relationship can be satisfactorily described as T$_{\rm N} = ({\rm M}_*/10^{6.7})^{-0.56}$ when M$_*$ is expressed in solar masses. The functional form of the relationship is only weakly constrained and extrapolation outside this range is not advised. Our late-type galaxies, in contrast to our early-types, do not show the tendency for low mass galaxies to split into two T$_{\rm N}$ families. Using these results and a galaxy stellar mass function from the literature, we calculate that in a volume limited, local Universe sample, clusters are most likely to be found around fairly massive galaxies (M$_* \sim 10^{10.8}$ M$_\odot$) and present a fitting function for the volume number density of clusters as a function of parent galaxy stellar mass. We find no correlation between T$_{\rm N}$ and large-scale environment, but do find a tendency for galaxies of fixed M$_*$ to have larger T$_{\rm N}$ if they have converted a larger proportion of their baryons into stars.
\end{abstract}

\keywords{galaxies: evolution, formation, star clusters, stellar content}

\section{Introduction}
\label{sec:intro}
In an effort to trace the properties of globular cluster populations, and by doing so constrain the star formation and merger histories of galaxies, we are undertaking a systematic, simple census of cluster populations in nearby galaxies using homogeneous and deep near-infrared imaging from the {\sl Spitzer Space Telescope} warm-mission survey of nearby galaxies \citep[S$^4$G;][]{sheth}. In the first paper in our series analyzing these data \citep[][herafter Paper I]{z15}, we presented results based on a set of 97 early-type galaxies. Here, we present results based on 73 edge-on late type galaxies, with the aim of confirming differences and similarities between the properties of the cluster populations that surround these very different morphological classes of galaxies and providing global parameterizations of the cluster population.

Because there is extensive literature regarding globular cluster populations, it is important to place this study in context. 
Compiling large, homogeneous samples of globular cluster population measurements has been challenging. In general, for comparative purposes, the number of clusters per galaxy is normalized, originally by parent galaxy luminosity but now more typically by parent galaxy stellar or total mass, and referred to as the globular cluster specific frequency \citep[see][for reviews]{harris,brodie}. 
The most recent and comprehensive compilation of specific frequencies is that of \cite{harris13}, who combed the literature to obtain specific frequency measurements for 422 galaxies.
However, this is a compilation from a myriad of existing studies, therefore,
despite careful efforts to homogenize the sample, some unquantifiable level of heterogeneity in sample selection, image quality and characteristics, image analysis, and cluster population modeling for sample completeness corrections is unavoidable. Because individual studies have tended to focus on a particular class of galaxies or environment, cross comparisons among galaxy types are susceptible to systematics.  A large, single data source study, at the very least, constitutes a complementary approach with which to address such concerns.

A second relative advantage of the sample described here is the set of large, uniform ancillary data and high level products already available for the S$^4$G galaxies. A number of studies of the S$^4$G data themselves exist and provide uniformly measured morphologies \citep{buta15}, photometry \citep{munoz}, stellar mass distributions \citep{querejeta}, and so forth. In addition, there are also now compilations of multiwavelength data for the S$^4$G galaxies ranging from the UV \citep{bouquin}, through the optical \citep{knapen}, to the radio \citep{courtois11,courtois15}. These data provide an advantage for S$^4$G cluster studies relative to literature compilations where not only are the specific frequency measurements heterogeneous but in many cases the ancillary data are either unavailable or themselves a compilation from multiple sources.

As there are advantages of a large study such as this, there are also disadvantages. Many recent studies, particularly those using high spatial resolution images obtained with the {\sl Hubble Space Telescope}, have aimed for high purity samples, where contamination among the candidate clusters is low \citep[see for examples,][]{peng06,strader,kundua,kundub,rhode04,young}. Our study does not have this level of purity and we cannot claim that any specific candidate cluster is truly a cluster. However, except for {a few studies \cite[such as those by][]{peng06,strader,villegas}, each of these high purity studies cover at most a few tens of galaxies because of the high observational cost. The Peng, Strader, and Villegas studies comprise larger samples}, but of galaxies within a cluster environment, which enables observations of large samples with such an oversubscribed facility as {\sl HST}. High purity samples are essential for detailed population or follow-up studies, particularly those requiring additional large telescope commitment, for example as required for spectroscopy \citep{sluggs}. As such, our study here is not competitive for those purposes.

For each galaxy, we present a measurement of the number of globular clusters normalized by parent galaxy stellar mass, T$_{\rm N}$, as advocated by \cite{zepf}. This basic quantity reflects the integrated efficiency with which a galaxy has formed and retained its cluster population. 
Following Paper I, where we developed our methodology and applied it to a sample of early type galaxies, 
we now measure the cluster populations of late type galaxies by quantifying the statistical excess of point sources surrounding galaxies in {\sl Spitzer Space Telescope} 3.6 and 4.5$\mu$m images from the S$^4$G survey \citep{sheth}. We treat the clusters in each galaxy as a single population, despite evidence from colors 
\citep{zepf,ostrov,gebhardt,larsen,kundua} and kinematics \citep{strader,woodley,pota} that they are not. As such, our measurements simply reflect the global population characteristics.
By relaxing the sample purity criterion, we accept greater uncertainties in exchange for a larger galaxy census that spans a broader range of galaxy properties. 
Ultimately, the final answer to whether lower precision measurements for a larger sample are scientifically useful depends on the magnitude of the differences present and the size of the sample. 
We have established that the methodology presented here is sufficiently accurate and precise to be scientifically useful in Paper I.
In \S 2 we describe the sample, how we construct the cluster candidate catalog, and how we measure the cluster specific frequency. We discuss our findings in \S3 regarding trends in T$_{\rm N}$ with galaxy properties, and conclude in \S4. 

\section{From Data to T$_{\rm N}$}
\label{sec:data}

\subsection{Constructing the Point Source Catalog and Radial Surface Density Profiles}

As in Paper I, the parent sample is the S$^4$G sample, which currently consists of 2,352 galaxies \citep{sheth}. It is primarily a volume-limited sample, but additional selection criteria,  such as the existence of an H{\small I} redshift preclude it from being a complete, volume-limited sample. We provide and analyze images of these galaxies obtained with the {\sl Spitzer Space Telescope} \citep{werner} and its Infrared Array Camera \citep[IRAC][]{irac}. The data are publicly available through the archive on the NASA IRSA website\footnote{http://irsa.ipac.caltech.edu/data/SPITZER/S4G/}. 

To complement the early-type sample of Paper I ($-5 \le$ T-type $\le 1$), we focus now on galaxies morphologically classified as late type ($4 \le$ T-type $ \le 10$) by \cite{buta15}, and select only those that are nearly edge-on (inclination $\ge 80^\circ$) to minimize the area on the sky in which 
internal structure in the parent galaxy could be mistaken for clusters and maximize the area in which the clusters will be detected and not
hidden by the parent galaxy. The inclinations are based on ellipticities measured in the S$^4$G pipeline \citep{munoz}. The inclination cut in particular results in a sample that is a small fraction of the full sample. We further constrain the sample by including only those galaxies within a suitable range of distances, again following the criteria defined in Paper I. We set the upper end of the distance range to correspond to a distance modulus of 32.4 (30.1 Mpc), where we have enough physical resolution to adequately sample the globular cluster population radial profile. We set the lower end of the distance modulus range at 30.25 (11.2 Mpc), where we ensure sufficient background coverage within the images with which to constrain the background source density. Finally, we remove from the sample any galaxies that have bright nearby neighbors that would compromise the analysis and the one galaxy (UGC 1839) in the S$^4$G sample that satisfies all of these criteria but for which there is not a well determined magnitude in the S$^4$G catalog. Because of these restrictive constraints, we re-examined all edge-on galaxies in the full sample for additional candidates. We found two (IC 5269 and NGC 1145) that are classified as early types (T-Type $-1$ and 0, respectively) but appear to be disk dominated and satisfy our distance and isolation criteria. We add these two galaxies to our sample and list the final 73 galaxies 
in Table \ref{tab:results}. 
We use redshift independent distances when available in NED, otherwise we use the redshift and adopt H$_0 = 72$ km s$^{-1}$ Mpc$^{-1}$ to derive a Hubble 
velocity distance estimate.

\begin{figure}[t]
\plottwo{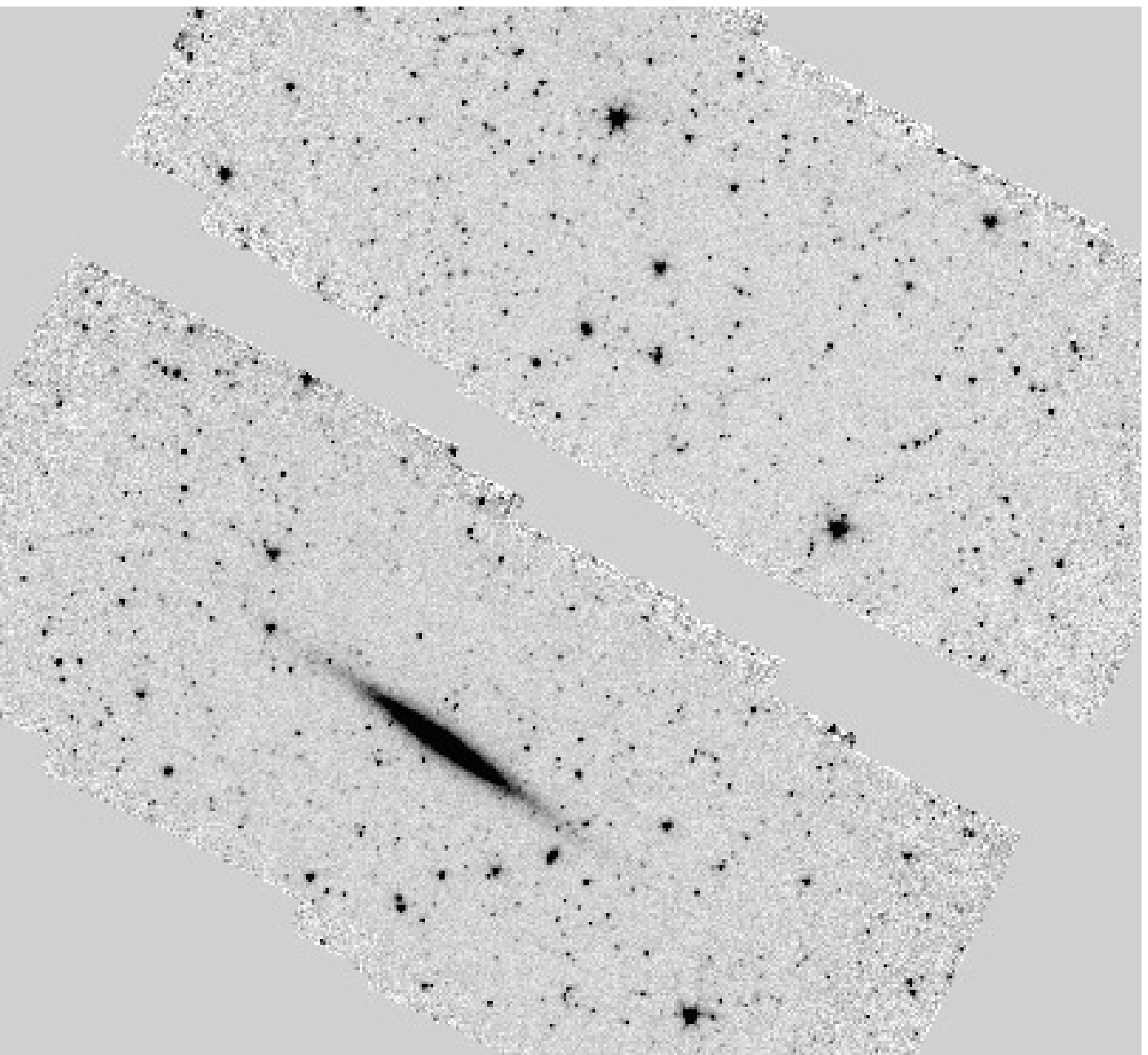}{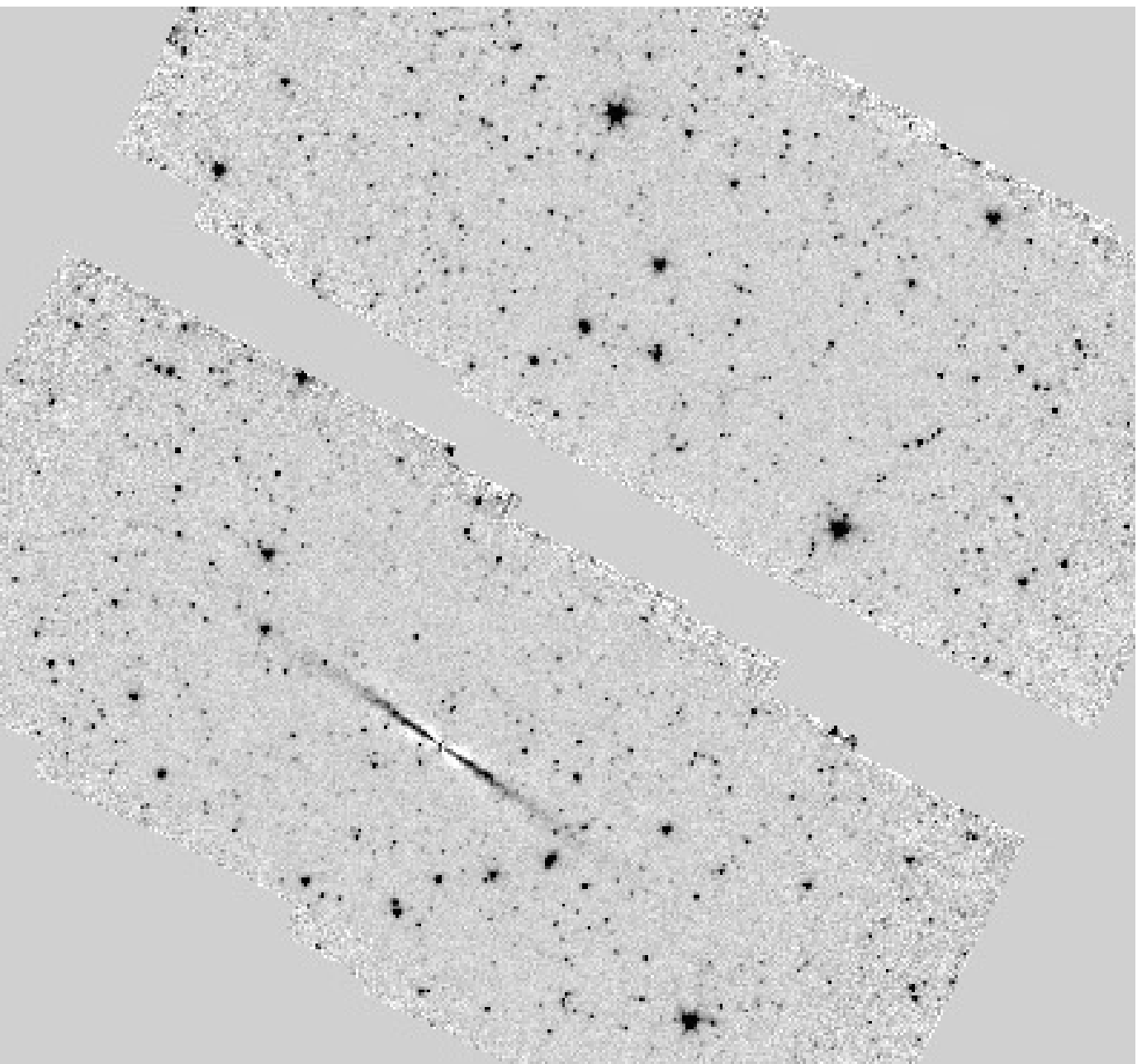}
\caption{Demonstration of the model subtraction for NGC 100. The results are not unusual. This is the S$^4$G produced mosaic in the 3.6$\mu$m passband. The angular size of this particular image is 12 arcmin across, corresponding to 57 kpc, and north is up.}
\label{fig:modelsub}
\end{figure}

\begin{figure}[t]
\epsscale{0.7}
\plotone{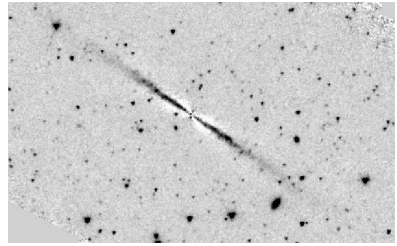}
\caption{Demonstration of the model subtraction near the core of NGC 100.  The vertical size of the image is about 3.5 arcmin or 16 kpc. The quality of the model subtraction is highly variable within a few kpc of the galaxy center along the major axis. Otherwise, sources can been detected to small galactocentric radii.}
\label{fig:modelzoom}
\end{figure}

To produce our measurements of the cluster population size, we follow the procedure described in detail in Paper I and briefly summarize here. We use the bright object masks and exposure weight maps
developed as part of the S$^4$G processing  \citep{salo,munoz}. 
We use the exposure weight maps to exclude areas with substantially less exposure time, and therefore lower sensitivity. The exact value of the thresholding we use varies for each image but is selected to exclude the image edges. Problems with detections at the image edges are often noticeable as sharp rises or dips in the final radial density profiles of sources and occur either at image gaps, for cases where multiple images are used to cover the field around a galaxy, or at the largest radii. We adopt the smallest threshold value that eliminates such features.  
We perform basic pre-processing steps including sky subtraction plus modeling and removal of the primary galaxy.
We calculate the background sky value by evaluating the median of the unmasked pixels within either the upper or lower quarter of the image, depending on whether the primary galaxy lies in one or the other of these two regions. We subtract this median sky value from the entire image. We then use the IRAF$^{}$ task ELLIPSE to measure the properties of the central galaxy, create an image of that model using BMODEL and then, by subtraction, obtain an image that is as nearly free of the primary galaxy as possible. 
Examples of the galaxy subtraction both on the scale of the full image and expanded about the target galaxy are shown in Figures \ref{fig:modelsub} and \ref{fig:modelzoom}, respectively, for NGC 100 (simply the first NGC galaxy on our list). The multipolar residual pattern seen in Figure \ref{fig:modelzoom} is typical. Except along the disk major axis, we can identify individual sources quite close to the center.
Once the residual images are available, we run SExtractor \citep{bertin} on unmasked versions to identify point sources, eventually using the stellarity index to reject clearly extended sources. For the inner regions, where the model subtraction is most important, we require consistent detection and photometry of candidate clusters in the 3.6 and 4.5$\mu$m images to help reject spurious sources. We apply this criterion only in the central regions because the two images generally do not overlap at large radii.
Candidate clusters are defined as objects within an absolute magnitude range given by $-11 < M_{3.6} < -8$, defined at the upper end by the known globular cluster luminosity function and at the lower end by requiring high completeness over our entire sample. We apply source count corrections as a function of both magnitude and location within the image, determined by adding artificial point sources over a range of magnitudes that is greater than that defined for the candidate clusters (because measurement uncertainties could move objects within our magnitude limits).  
Details of parameter choices in this and previous steps are given in Paper I. We made no modifications to the procedure between this paper and our previous work and present validation of these parameter choices in Paper I. 

\subsection{Parametrizing the Cluster Population}

Using the radially binned, completeness corrected surface source density values, we now estimate the parameters of a power-law profile description of the cluster distribution. The data are of insufficient quality to allow for the fitting of the power law and background simultaneously. We follow the preferred approach from Paper I, where we fix the power-law slope at $-2.4$ and vary the power-law normalization and background level, still fitting the model for radii between 1 and 15 kpc by minimizing $\chi^2$.  The power-law exponent was set in Paper I by maximizing the concordance between our results and those of \cite{harris13} for galaxies in common. The specific radial range was set at the lower end by the minimum radius at which the model subtraction provided low enough residuals for cluster detection and at the upper end by the maximum radius at which there was often signal above background in the cluster surface density profiles. 
As shown in Figure \ref{fig:profiles}, the power-law plus constant background model does a satisfactory job of fitting almost all 73 galaxies.

We present the adopted and resulting values for the distance modulus (DM), Hubble T-Type (T), the 3.6 and 4.5$\mu$m apparent magnitudes, integrated number of clusters { from a radius of 5 kpc out to a radius of 50 kpc} (N$_{50}$), the specific frequency relative to the galaxy's stellar mass (T$_{\rm N}$) corresponding to N$_{50}$ in units of number per 10$^9$ M$_\odot$ (as introduced by \cite{zepf}), the logarithm of the surface number density of background sources, and quality flag (Q, described below) in Table \ref{tab:results}. 
{The choice of the inner boundary in the integration affects the total numbers of clusters inferred, but not the relative numbers among galaxies because of our choice of a fixed power law slope. The specific choice of a 5 kpc integration lower limit is inherited from Paper I, where the sample consists of large elliptical galaxies. Perhaps a more appropriate choice now is a radius of 3.1 kpc, which is the smallest radius for which we have measurements in all of the galaxies we will consider (see below). Changing the lower bound of the integration from 5 to 3.1 kpc results in an increase of all the globular cluster specific frequencies by 35\%, but we retain the 5 kpc inner bound for consistency with Paper I. We will return to this issue when we discuss globular cluster specific frequencies in an absolute context}.

\begin{figure*}
\epsscale{1}
\plotone{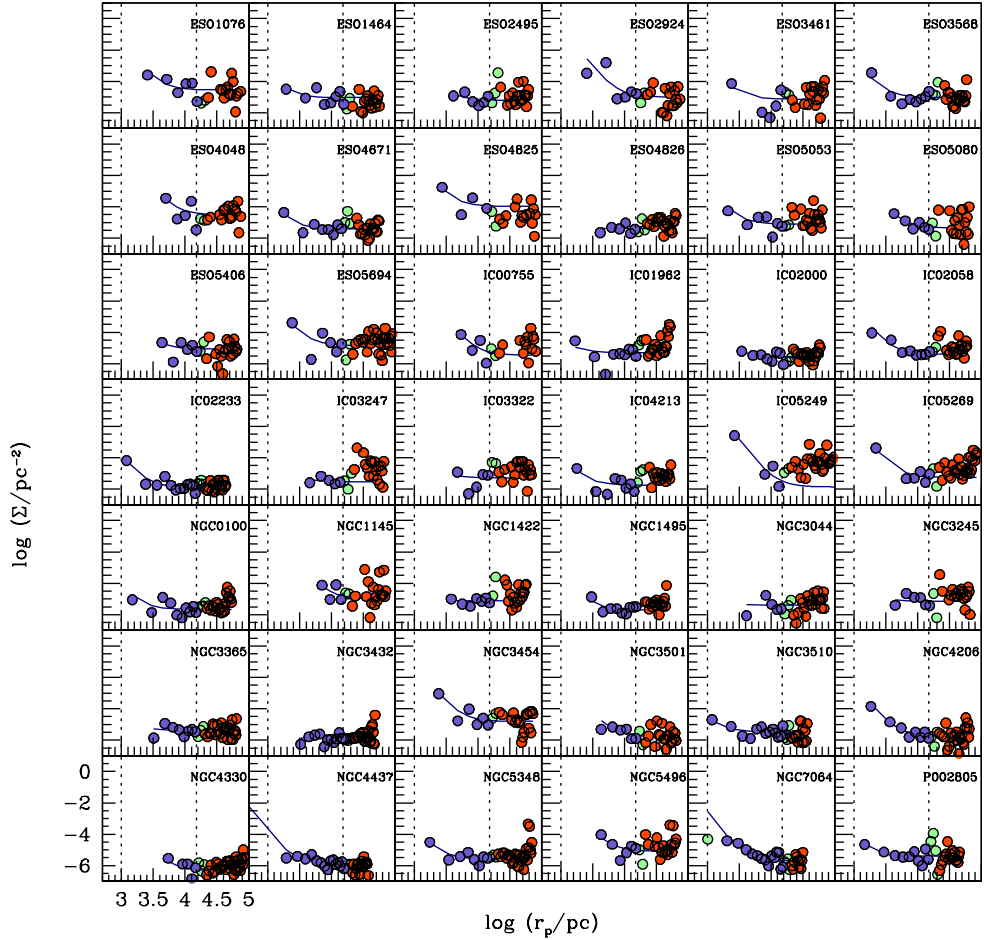}
\caption{Surface number density radial profiles of candidate globular cluster populations. Each panel contains the data for one galaxy.  The two vertical dotted lines denote the radial range over which the power-law model is fit (1 to 15 kpc). Data within that range are plotted as blue circles. The red circles denote the data used to determine the background source level and includes all data beyond 30 kpc. Data that are neither in the fitting range or background range are plotted as light green. The solid line shows the best fit model plus background over the radial range for which data exist. In the case of NGC 4437, where the solid curve appears to extend inward farther than the data, there is one datum beyond the lower limit of the plot and the inward extrapolation of the model is consistent with it.} 
\label{fig:profiles}
\end{figure*} 

\setcounter{figure}{2}
\begin{figure*}
\plotone{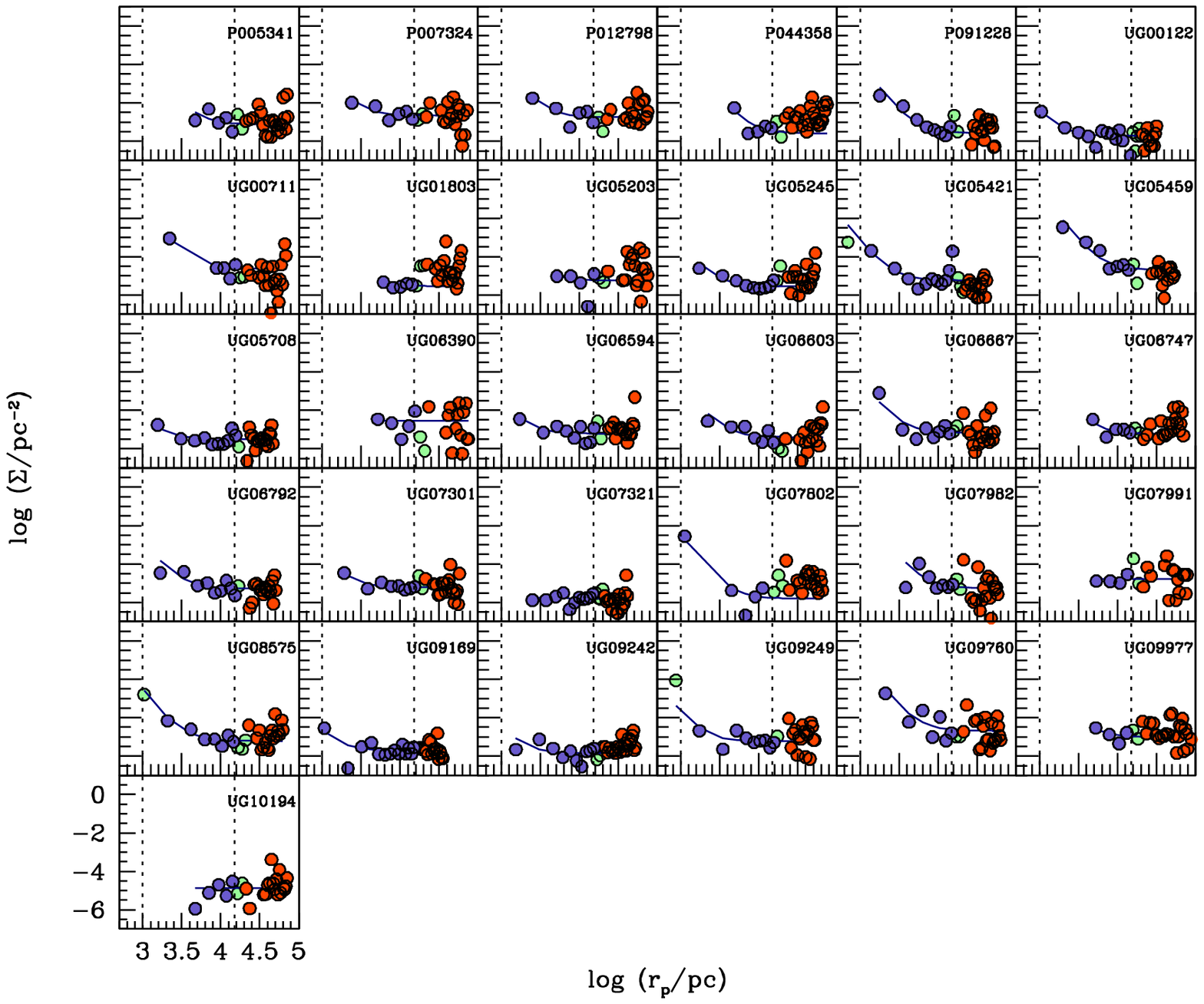}
\caption{cont.}
\end{figure*} 

We base our uncertainty estimates for T$_{\rm N}$ on Poisson statistics in the individual radial bins, propagated through the fitting using $\Delta \chi^2$. In cases where the model fit is statistically acceptable (defined as having a probability $>$ 0.33 of the data being drawn from model) we adopt the uncertainties corresponding to the $1 \sigma$ range in the model parameters. In cases where the model is statistically unacceptable, for the adopted Poisson uncertainties in the individual bins, we calculate the value of $\chi^2$ at which the probability of the data being drawn from the model is 0.5. We increase the uncertainties in each bin, all by the same scaling factor, so that the resulting $\chi^2$ now has this value. Rescaling is required for the majority of the galaxies because the internal (Poisson) statistics are often underestimates of the full uncertainty. This approach corresponds to using the scatter about the fit to estimate the true uncertainty. In Paper I we externally validated the determined uncertainties by comparing our measurements of the number of clusters, N$_{\rm CL}$, to those in the literature for the subset of galaxies for which this comparison is possible. Unfortunately, we cannot expand on that comparison here because none of the galaxies in this study are also in the \cite{harris13} study.

We noted in Paper I that  $\chi^2$ only judges models where data exist, but that data may not exist over the critical range of parameter space. We then quantified 
how well the data constrain the model over the key radial range of 1 to 15 kpc by applying a binary flag, $Q = 0$ for galaxies in which the data do not reach interior to $\log r = 3.5$ (3.1 kpc) and 1 otherwise. 
This quality index for our new sample of galaxies is included in Table \ref{tab:results}. In Paper I and here, we opt to use only $Q=1$ profiles for our subsequent discussion and, by doing so, remove 33 galaxies from our sample for subsequent discussion. This choice was supported in Paper I by the superior matches between our measurements and those of \cite{harris13} for the $Q=1$ sample.

As in Paper I, we quote the integrated number of clusters within 50 kpc based on the best fit profile. The question of how far out in radius to integrate the profiles had greater bearing in Paper I because there we explored fitting different power law slopes to different galaxies. Ultimately, we decided that the data were insufficiently constraining to allow this freedom and settled on the fixed power law slope. Because we adopt that constant power law slope here, the decision to limit our cluster counts to radii $\le$ 50 kpc does not affect the relative number of clusters we measure among galaxies.
We integrate the power-law profile to $r = 50$ kpc and then correct that number for clusters outside of the magnitude range of our detected candidate clusters assuming a Gaussian luminosity function that is the same for all galaxies. We adopt the same standard parameters for the peak and width of the luminosity function as in Paper I; M$_V \sim -7.4$ and $\sigma_V = 1.4$ for early types and $\sigma_V =$ 1.2 for later types \citep{brodie}. For a V$-$3.6 color of $\sim$2.4 \citep{barmby} the location of the LF peak lies at M$_{3.6} = -9.8$. The dispersion of the cluster LF is not well measured at 3.6$\mu$m so we adopt the lower range of $\sigma$ estimates in the $V$ band, $\sigma_{3.6} = 1.2$.  There is little variation in the globular cluster luminosity function with galaxy luminosity \citep{strader}. Variations of these parameters, within reason, tend to change in the numbers of clusters by tens of percent, rather than by factors of a few, which is what we concluded in Paper I to be the actual uncertainty of our measurements. Nevertheless, one of the greatest sources of systematic uncertainty comes from our adoption of universality in the globular cluster population radial profile shape and luminosity function. Comparison to external studies (see Paper I), that make different assumptions provides guidance on the magnitude of this uncertainty and supports our adopted uncertainties.

To convert the number of clusters to a specific frequency we use the stellar mass of the parent galaxy as calculated from the {\sl Spitzer} magnitudes and their calibration to stellar mass \citep{eskew}. { More detailed stellar mass and stellar population modeling of the S$^4$G galaxies exists in a set of studies \citep{meidt12,meidt14,querejeta,rock}}, but using the \cite{eskew} calibration, which is coarser but consistent with those other studies, provides a direct, self-consistent mass estimate for the clusters and enables straightforward extension to other galaxies beyond those in S$^4$G. 

\section{Discussion}

\begin{figure}
\epsscale{0.9}
\plotone{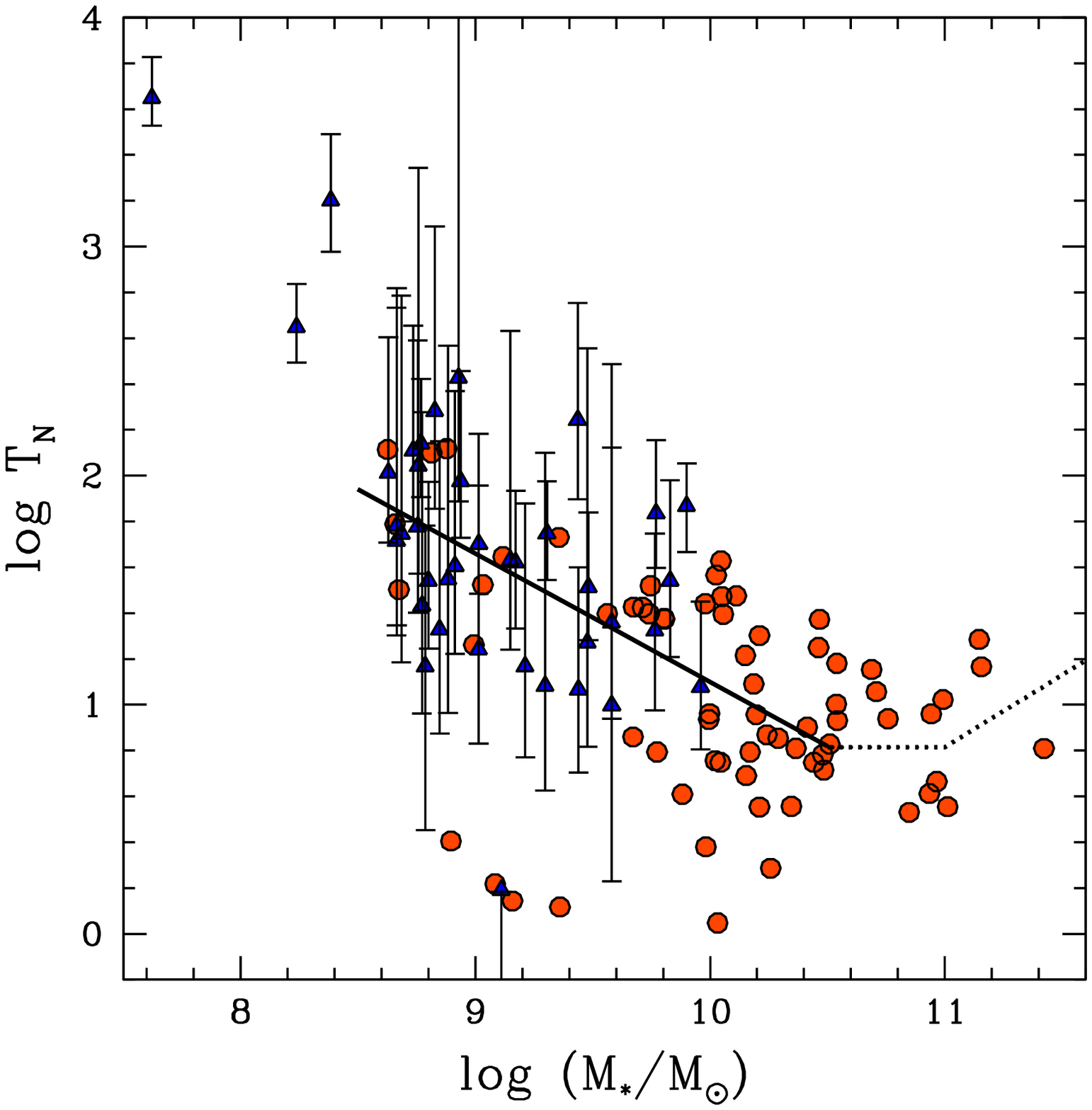}
\caption{Cluster population stellar mass normalized specific frequency (T$_{\rm N}$) versus parent galaxy stellar mass. Red circles denote early type galaxies from Paper I  (T $\le 1)$ and blue triangles denote the current sample of edge-on late type galaxies ($4 \le {\rm T} \le  10)$. Only galaxies with a quality flag of 1 and with photometry in both IR bands are included. The solid line indicates the unweighted linear fit for $8.5 < \log {\rm M}_*/{\rm M}_\odot < 10.5$, where we have most of our data. The dotted line represents the possible behavior as suggested by our data and the \cite{peng08} results for the most massive galaxies (see text for details). The uncertainties on the early type galaxy measurements are comparable to those of the late-type galaxies, but they are omitted for clarity.}
\label{fig:specfreq}
\end{figure}

As we found in Paper  I and also see in Figure \ref{fig:profiles}, a large fraction of galaxies exhibit a radially clustered unresolved source population, which presumably consists of globular clusters. A power-law plus constant background model is an adequate description for the radial distribution of these sources given the current state of the observations for most galaxies. Proceeding along the lines of Paper I so that we can compare to the results for early-type galaxies, 
we show the stellar mass normalized specific frequency of clusters, T$_{\rm N}$, as a function of parent galaxy stellar mass, M$_*$, in Figure \ref{fig:specfreq} for both the current late-type sample and our previous early-type sample. 

Some basic qualitative results can be drawn quickly from Figure \ref{fig:specfreq}: 1) there is a general decrease in T$_{\rm N}$ as stellar mass rises, 2) despite this mean trend, for M$_* < 10^{10}$ M$_\odot$ the variation in T$_{\rm N}$ can be larger than a factor of 10 at a given M$_*$, 3) the fraction of the population with extremely low T$_{\rm N}$ (log T$_{\rm N} <  0.5$) at low M$_*$ (M$_* < 10^{9.5} {\rm M}_\odot$) is smaller among late type galaxies, where there is only one such galaxy (UGC 7321) out of 41 galaxies (0.024), than among similar massive early type galaxies, where there are 4 out of 23 (0.17), 4) for galaxies with M$_* > 10^{10.5} {\rm M}_\odot$ the behavior of T$_{\rm N}$ can be characterized as flat or at least flatter (although this result is not impacted by the current sample, which adds no galaxies with M$_* > 10^{10.5}$ M$_\odot$ to the combined sample), and 5) for galaxies with M$_* < 10^{8.5} {\rm M}_\odot$ the rise in T$_{\rm N}$ appears to accelerate with decreasing mass, although we have only three galaxies in this regime and the potential for systematic problems in both the mass estimates and the cluster counts is large.

Before discussing  these results in more detail, we consider two potential problems with expressing the measurements as done Figure \ref{fig:specfreq}. First, because both axes in the plot depend on M$_*$, errors in M$_*$ may produce apparent trends. For example, underestimating the true stellar mass, moving that particular galaxy leftward in the Figure, would also result in overestimating T$_{\rm N}$, moving that galaxy upward in the Figure. The sense of such correlated errors is therefore the same as that of the observed trend. However, this ambiguity is resolved by observing galaxies over a much larger range of M$_*$ than can be accounted for by errors in M$_*$.  Over the best sampled region of the Figure, we span a factor of 100 in stellar mass, $8.5 < \log({\rm M}_*/{\rm M}_\odot) < 10.5$, a range that is much larger than the size of the uncertainty in M$_*$ \citep[30\%;][]{eskew}. We conclude that errors in M$_*$ are not responsible for the apparent correlation. Second, because the sample is composed of galaxies of different populations, apparent trends might arise if T$_{\rm N}$ and M$_*$ vary grossly among populations, but not necessarily in a correlated manner within each population. For example, if late-type galaxies are both less massive and have larger T$_{\rm N}$ than early type galaxies, then placing the two populations of galaxies on the sample plot would result in an apparent correlation between T$_{\rm N}$ and stellar mass, even if there is no such correlation within each population. This concern is ameliorated both by noting that each population independently shows the correlation and that there is substantial overlap along the abscissa of Figure \ref{fig:specfreq} for the two populations. 

In light of these considerations, we conclude that we confirm the finding by \cite{harris13} that among galaxies with log (M$_*/{\rm M}_\odot) < 10.5$ there is an inverse correlation between T$_{\rm N}$ and M$_*$. An unweighted linear fit to the data for galaxies with $8.5 < \log M/M_\odot < 10.5$ yields the solid line shown in Figure \ref{fig:specfreq}, which can be expressed as ${\rm T}_{\rm N} =  10^{6.7} (M_*/M_\odot)^{-0.56}$. In our subsequent search for potential sources of physical scatter in T$_{\rm N}$, we will use this relationship as a fiducial against which to search for correlations between other characteristics and deviations of T$_{\rm N}$ from the mean. Although we conclude that there is a correlation between
T$_{\rm N}$ and M$_*$, we caution that, as always, correlations do not imply causality and that the existing data suggest that this relationship does not extend beyond the quoted stellar mass range.

The situation at small M$_*$  (M$_* < 10^{9.5}$) is a bit muddled. Most studies, including ours, find significant scatter in specific frequency among low mass galaxies \citep[for example, luminosity normalized specific frequency values, S$_{\rm N}$, in one sample range from 0 to 23 for a set of dwarf elliptical galaxies;][]{miller}. \cite{strader} suggest that there are two families of galaxies among the dwarf ellipticals, one with S$_N \sim 2$ and another with S$_N \sim 5 - 20$. Although our numerical values of specific frequencies are not directly comparable to the literature values because of the use of different filter bands and our conversion to stellar mass, our results for early type galaxies in Paper I were consistent with the suggestion of two families. On the other hand, we find little evidence for two populations in our new late-type galaxy sample, which has twice as many galaxies in the relevant stellar mass range and only one galaxy that falls on the lower T$_{\rm N}$ branch as defined by the early type galaxies. The remainder of our late type galaxies fall on the upper branch. 
The existence of two populations is also not evident in the compilation of \cite{harris13}. 
 
Our data show a marked change in the behavior of T$_{\rm N}$ with M$_*$ for  M$_* > 10^{10.5}$ M$_\odot$, consistent with the results of \cite{georgiev} and \cite{harris13}. However, the new sample presented here does not influence this result because the late-type galaxy sample has no systems with masses above this stellar mass cutoff, it merely strengthens the case for a trend of T$_{\rm N}$ with M$_*$ at lower masses. At even higher masses, $M_* > 10^{11}$ M$_\odot$, \cite{peng08} find that T$_{\rm N}$ reverses and begins to rise again. We
have illustrated the range of behavior with our three part function in Figure \ref{fig:specfreq}, where the rising part is defined to have a slope derived from the \cite{peng08} data for
M$_* > 10^{10.5}$ M$_\odot$ but is only applied here for $M_* > 10^{11}$ M$_\odot$. Our data are consistent with this parameterization, but offer little in the way of a constraint at these masses. The \cite{peng08} data is for ellipticals in the Virgo cluster, galaxies that may have experienced a different history and thereby developed a different globular cluster population than field galaxies at these masses. We therefore suggest caution in interpreting this ``spliced" functional form for T$_{\rm N}$ even though it is consistent with our data.

\begin{figure}
\epsscale{0.9}
\plotone{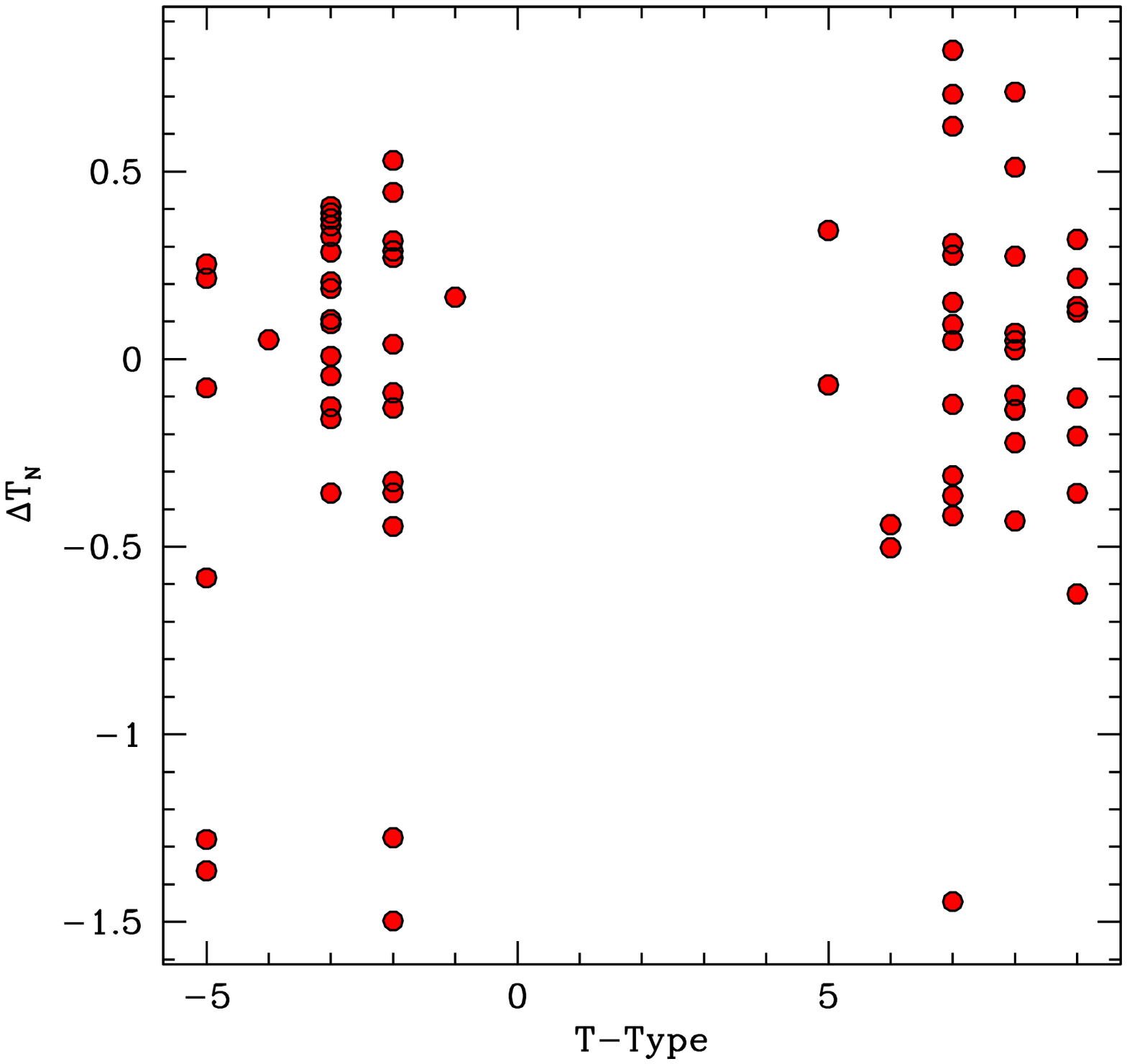}
\caption{Residuals from the mean trend of T$_{\rm N}$, $\Delta {\rm T}_{\rm N}$, vs. morphology (T-Type) for galaxies with $8.5 < \log ({\rm M}_*/{\rm M}_\odot) < 10.5$. }
\label{fig:morph}
\end{figure}

We confirm and extend the finding from Paper I that 
morphology plays at most a minor role in determining T$_{\rm N}$. We find no evidence for a difference in T$_{\rm N}$ as a function of morphology, other than the possible difference in how the high and low T$_{\rm N}$ branches at low M$_*$ are populated (Figure \ref{fig:morph}). We note, however, that care must be taken when comparing the properties of cluster populations around parents of different morphology. Because late-type galaxies are generally of lower stellar mass than early type galaxies of similar luminosity, and because of the relationship between T$_{\rm N}$ and M$_*$, a straightforward comparison of cluster specific frequencies of early and late type will observe  a difference.
Of course, our uncertainties are large and modest differences among galaxy populations could be masked by those uncertainties. Larger samples, which are possible within S$^4$G if we relax some of the selection criteria, would help address this issue even if the single galaxy T$_{\rm N}$ uncertainty remains the same. Similarly, we find no significant correlation between T$_{\rm N}$ and whether a galaxy is barred or unbarred.

\subsection{How Efficient is Cluster Formation and Where are Most Clusters Today?}

We use the mean trend between T$_{\rm N}$ and M$_*$ to address two simple questions. What fraction of stars end up in long-lived, massive clusters? Where are most clusters found in the current Universe? Using the results we present in Figure \ref{fig:specfreq} we adopt the following expression for T$_{\rm N}$:

\begin{equation}
{\rm T}_{\rm N} = \begin{cases}
10^{6.7}{\rm M}_*^{-0.56}&{\rm if\ } 10^{8.5} \le \ {\rm (M}_*/{\rm M}_\odot) < 10^{10.5} \\
8.3 & {\rm if \ }  10^{10.5} \le ({\rm M}_*/{\rm M_\odot }) \le 10^{11} \\
10^{-6.11}{\rm M}_*^{0.63} & {\rm if \ (M}_*/{\rm M}_\odot) > 10^{11} 
\end{cases}
\end{equation}

\noindent
which simply follows the fit shown in Figure \ref{fig:specfreq} for $8.5 < \log ({\rm M}_*/{\rm M}_\odot) < 10.5$, is flat for intermediate M$_*$, and then rises for the most massive galaxies as found by \cite{peng08}. Although we choose a T$_{\rm N}$ slope to match the \cite{peng08} data in this mass range, we normalize their relation to achieve continuity with ours at lower masses. For an estimated mean cluster mass of $1.2\times 10^5$ M$_{\odot}$, obtained from our adopted cluster luminosity function and the \cite{eskew} mass calibration, we express the cluster stellar mass fraction, the fraction of stars in a galaxy that are in clusters vs. the field, as $1.2\times 10^{-4}  {\rm T}_{\rm N}$, which when combined with the expression above for T$_{\rm N}$ suggests a moderately increasing cluster stellar mass fraction with decreasing parent galaxy mass over the stellar mass range that we best constrain, $8.5 < \log ({\rm M}_*/{\rm M}_\odot) < 10.5$.

The resulting values of the mass fraction range from $10^{-3}$ at M$_* = 10^{10.5}$ M$_\odot$ to 0.013 at M$_* = 10^{8.5}$ M$_\odot$, or roughly a factor of ten increase over the range of parent galaxy stellar mass explored best here. We discussed in Paper I how interpreting the cluster stellar mass fraction as representative of the cluster formation efficiency for lowest mass galaxies in this range matches theoretical expectations in some models for the cluster formation efficiency in dwarf galaxies at high redshift \citep{elmegreen12}, which further suggested that the cluster populations in these galaxies may be dynamically undisturbed to the present day. { We caution that global changes to the inferred cluster mass fractions will also arise with changes to the adopted inner integration boundary of the cluster radial distribution profiles used to calculate the total number of clusters in each galaxy. Any such change will affect all galaxies in equal proportions thereby leaving comparisons among galaxies unaffected.}

The decreasing cluster stellar mass fraction, as we move toward higher parent galaxy stellar masses, could reflect either a true decrease in cluster formation efficiency or a greater rate of cluster disruption. Theoretical models show that cluster destruction can be extremely effective in massive galaxies \citep{gnedin}, resulting in the destruction of as many as 90\% of the original clusters. If the destroyed fraction is this large at the upper end of our stellar mass range and if the rate of dynamical evolution decreases in lower mass galaxies, then dynamical evolution of the cluster populations would be a straightforward explanation of the observed trend. The problem, however,  involves a complex interplay of various factors and must be treated carefully. A recent study by \cite{gnedin14} that explores the interplay between the evolution of the globular cluster population and the growth of nuclear star clusters, and eventually central black holes, is one example of how such modeling can proceed. For higher parent galaxy stellar masses, $\sim 10^{11}$ M$_\odot$ and above, they predict an increasing fraction of stellar mass in globular clusters, consistent with the \cite{peng08} results that we have spliced onto Figure \ref{fig:specfreq}, unfortunately they do not model lower mass parent galaxies. { \cite{mieske} specifically show how the qualitative behavior we observe in $T_{\rm N}$ vs. M$_*$ can arise from tidal disruption.}

\begin{figure}
\epsscale{0.9}
\plotone{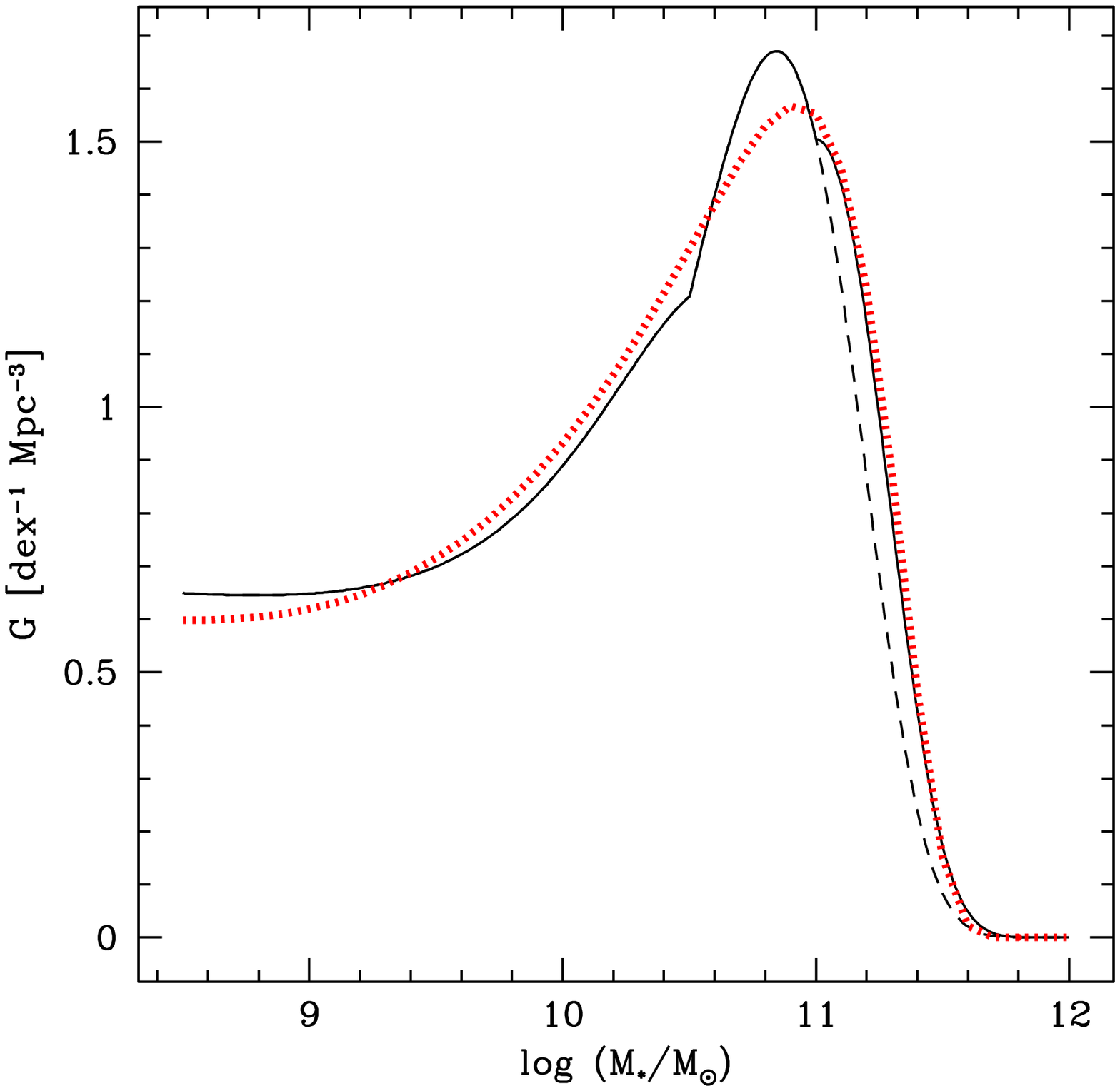}
\caption{Globular cluster distribution function as a function of parent galaxy stellar mass. The noticeable kinks in the function at log(M$_*/{\rm M}_\odot) = 10.5$ and 11 are due to the transition in derivatives among the segments in the T$_{\rm N}$ vs. M$_*$ relation (Eq. 1). The solid line represents the results adopting Eq. 1, while the dashed line the results if we adopt a constant T$_{\rm N}$ for log (M$_*/{\rm M}_\odot) > 10.5$. Details of the calculation are given in the text. The red dotted line represents our fitting function, Eq. \ref{eq:fit}.}
\label{fig:gcdist}
\end{figure}

Although the cluster stellar mass fraction, as defined by present day clusters, is highest in lower mass galaxies, addressing whether these galaxies also contain the bulk of clusters is  complicated by having to account for the relative numbers of galaxies of different stellar masses. The cluster distribution function among galaxies of different M$_*$ is given by the combination of T$_{\rm N}$ and $\phi({\rm M}_*)$, where $\phi({\rm M}_*)$ represents the volume density of galaxies with M$_*$. Taking parametrized expressions for each of these into account, adopting the stellar mass function measured from the GAMA survey by \cite{baldry}, we calculate the globular cluster distribution function shown in Figure \ref{fig:gcdist}. The Figure clearly shows that the bulk of today's clusters are found in galaxies with M$_* \sim 10^{10.8}$ M$_\odot$, which are relatively massive galaxies. The greater numbers of less massive galaxies combined with the somewhat larger T$_{\rm N}$ values for those galaxies was insufficient to counterbalance the fact that the more massive galaxies also have a larger absolute number of clusters. To enable further calculations with this globular cluster distribution function, G$({\rm M_*})$, we provide the following fitting function, which is also shown in Figure \ref{fig:gcdist}:
\begin{equation}
{\rm G}({\rm M_*}) = 0.9 \exp (-({\rm M}_*/10^{11.3})^{2.3}) ((\log {\rm M}_* - 8.4)^{2.8} + 6.64)
\label{eq:fit}
\end{equation}
The parameters (normalizations, exponents, and constants) in Eq. \ref{eq:fit} where determined by minimizing $\chi^2$ for the selected functional form. We did not explore a wide range of functional forms, so the equation is simply meant to be a convenient fitting function over the range of stellar masses plotted in Figure \ref{fig:gcdist} and we do not ascribe physical meaning to the function or the fitted parameters.

There are two caveats to this result. First, we do not measure the cluster populations of the most massive galaxies ourselves. We have adopted a functional form for T$_{\rm N}$ at the highest masses that is consistent with the \cite{peng08} data. While this approach may not be appropriate for galaxies outside of clusters, the most massive galaxies tend to be found mostly in clusters. Furthermore, as discussed in Paper I, the possibility that the stellar initial mass function varies systematically among early type galaxies could account for the entire observed rise in T$_{\rm N}$ at these masses. Fortunately for our calculation, these massive galaxies are exceedingly rare, so significant variations in the behavior of T$_{\rm N}$ in this mass regime have modest effects on the globular cluster distribution function. To demonstrate that the effects are minor, we recalculate G(M$_*$) adopting instead a constant T$_{\rm N}$ for M$_* > 10^{10.5}$ M$_\odot$ and show the difference with our previous calculation in Figure \ref{fig:gcdist}. Second, we have not extrapolated our T$_{\rm N}$ relation to M$_* < 10^{8.5}$ M$_\odot$ because we have little data at those masses. The \cite{baldry} galaxy stellar mass function also does not extend below M$_* = 10^8$ M$_\odot$.  However, if we extend the T$_{\rm N}$ relation down to M$_* = 10^8$ M$_\odot$, then we do find the cluster numbers rising at low M$_*$, although slowly. If T$_{\rm N}$ is grossly larger for low mass galaxies, as hinted at by the three galaxies in our sample at lower M$_*$, then a significant number of clusters could be hosted by such galaxies.

\subsection{Drivers of Variations in T$_{\rm N}$ at Fixed M$_*$}

The scatter in our measurements of T$_{\rm N}$ is large, although not manifestly larger than that plausibly attributable to measurement uncertainties (Figure \ref{fig:specfreq}). However, we might still be able to uncover physical sources of scatter in T$_{\rm N}$ if those contribute a comparably large level of apparent scatter. To investigate, we correlate various characteristics of our galaxies with deviations, $\Delta {\rm T}_{\rm N}$, from the mean relation between T$_{\rm N}$ and M$_*$, within the stellar mass range for which that relationship is best defined, $8.5 < \log ({\rm M}_*/{\rm M}_\odot) < 10.5$.  

\begin{figure}
\epsscale{0.9}
\plotone{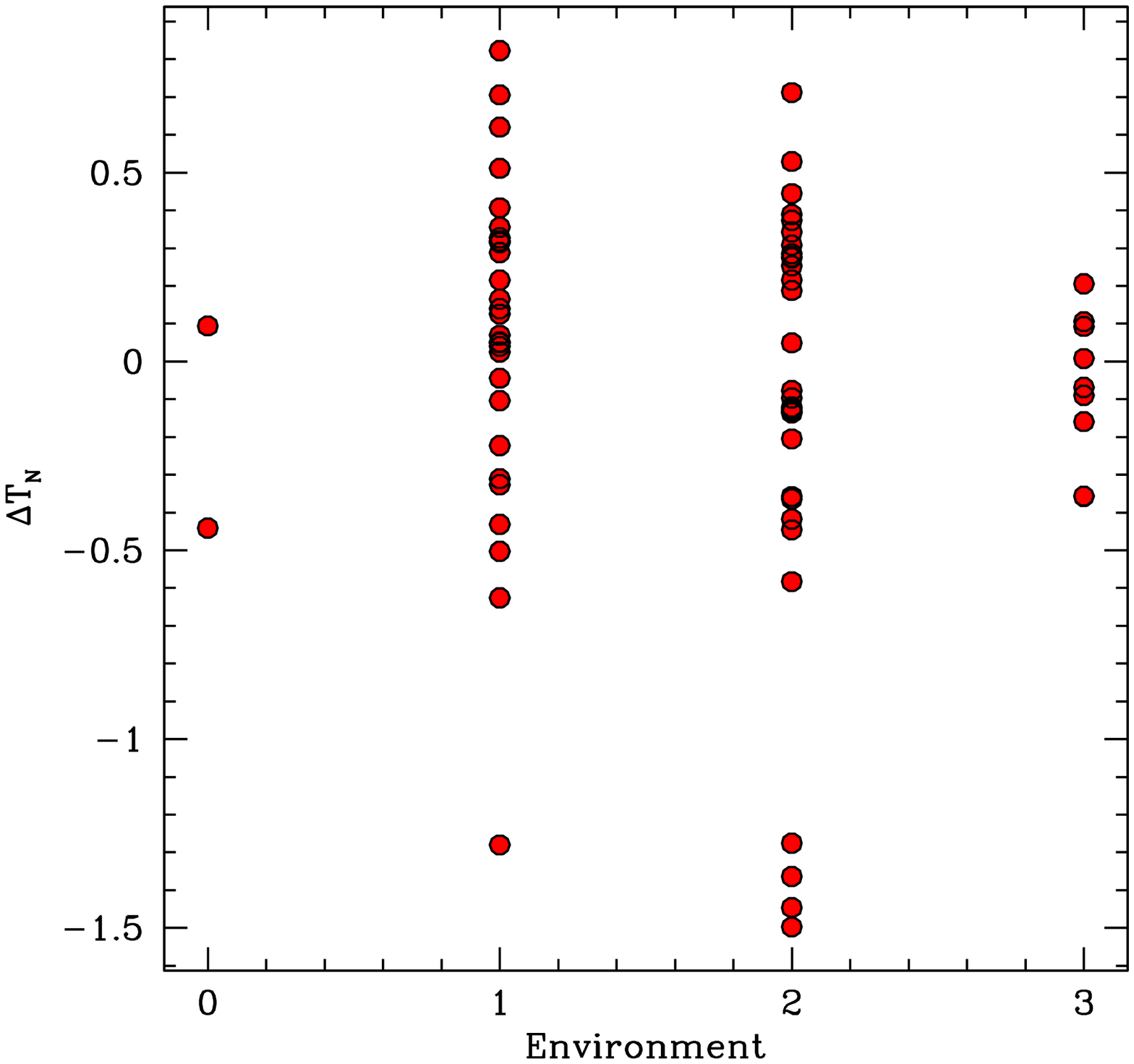}
\caption{The residuals about the mean trend for stellar mass normalized specific frequency, $\Delta {\rm T}_{\rm N}$, vs. M$_*$ plotted against our large scale environment measurement for galaxies with $8.5 < \log ({\rm M}_*/{\rm M}_\odot) < 10.5$. The environment measurement is obtained by measuring the velocity shear field at the position of the galaxy \citep{courtois15} and is numerically quantified as $0 = $ void, $1 = $ filament, $2 = $ sheet, and $3 = $ knot. We find no evidence for a correlation between specific frequency and this measurement of environment for intermediate mass galaxies.}
\label{fig:environment}
\end{figure}

The first characteristic we explore is the galaxy's large scale environment as measured using the cosmic shear field \citep{courtois15,courtois}. We have previously used this measurement to investigate the nature of extremely massive and gas rich galaxies within the S$^4$G sample \citep{courtois}. This environment measurement is a coarse one, indicating only whether the galaxy lies is a void, filament, sheet, or knot (designations 0,1,2, and 3, respectively). We find no significant correlation between $\Delta {\rm T}_{\rm N}$ and environment, although the scatter among T$_{\rm N}$ values appears to be smaller in the densest environments, the knots (Figure \ref{fig:environment}). This visual impression is supported by the results of a statistical F-Test to determine the likelihood that any two sets of data were drawn from a Gaussian distribution with the same variance. Comparing the filament and sheet galaxies with those in the knots results in probabilities that they were drawn from the same parent sample of 0.01 and  $7\times 10^{-4}$, respectively, suggesting that the variances of the underlying populations are actually different. The one strong caveat for this test is that it is highly sensitive to non-Gaussianity in the underlying distributions. This result merits attention with larger samples, but for now we conclude that we find no clear sign of large-scale environmental effects on the specific frequency of globular clusters, for galaxies in this intermediate mass range. 

\begin{figure*}
\epsscale{0.9}
\plotone{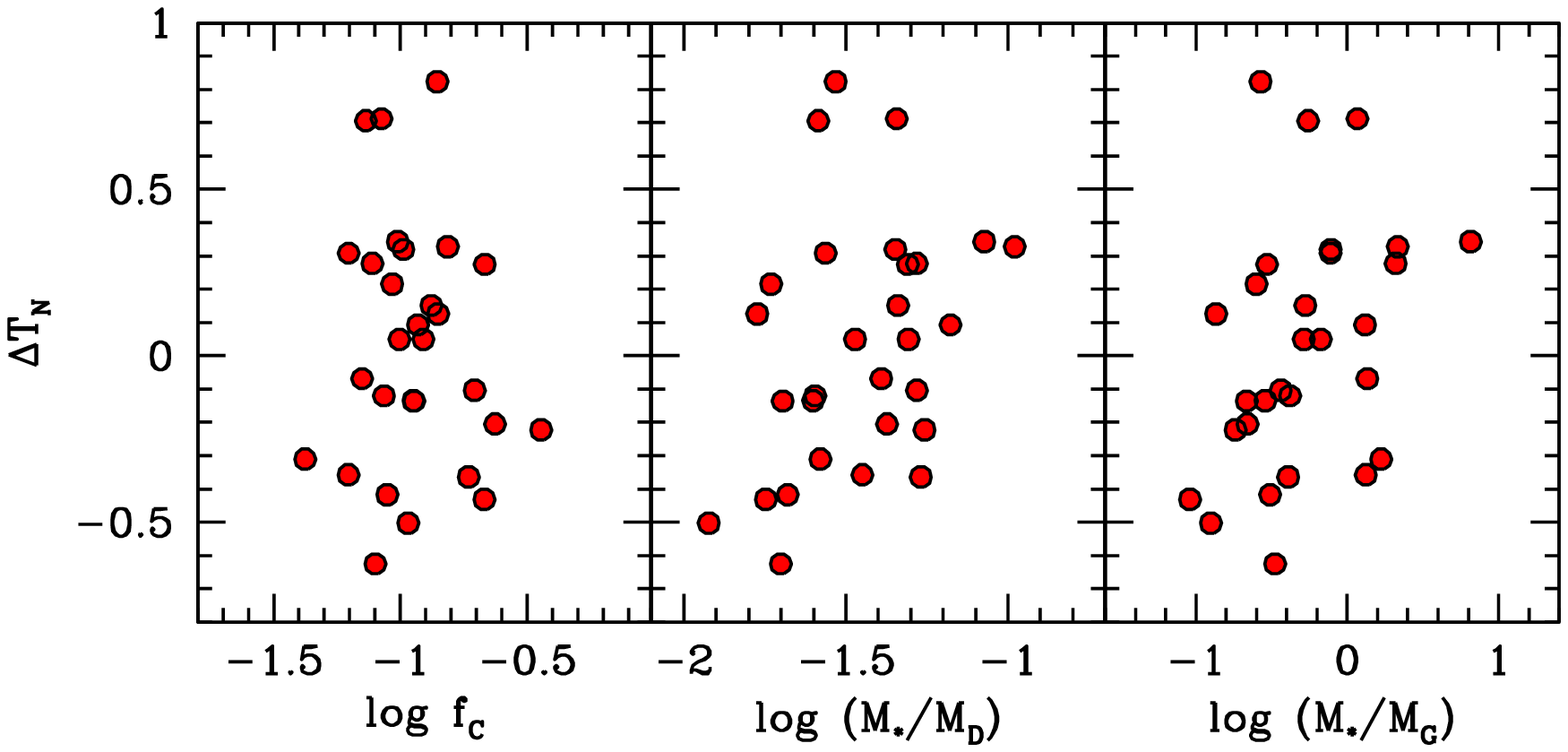}
\vskip -1in
\caption{The residuals about the mean trend for stellar mass normalized specific frequency, $\Delta {\rm T}_{\rm N}$, vs. various baryon-sensitive parameters. The leftmost panel plots the relationship between $\Delta {\rm T}_{\rm N}$ and the fraction of the cosmic baryon budget for each galaxies that is detected as stars or cold gas. This ratio is what we refer to as the condensed baryon fraction, $f_C$, and represents the relative baryonic mass that has settled into the luminous portion of the galaxy \citep{zbtf}. The middle panel shows $\Delta {\rm T}_{\rm N}$ vs. the ratio of mass in stars to that in the dark halo. Finally, the rightmost panel shows the ratio of mass in stars to that in cold gas. Each of these represents in different ways the efficiency with which baryons in a dark matter halo are concentrated toward the center of the dark matter halo and turned into stars.}
\label{fig:correlations}
\end{figure*}

Next we explore the relations between $\Delta {\rm T}_{\rm N}$ and measurements of the galaxies' baryonic and dark matter content. The rationale for such explorations rests with questions regarding the relative efficiency of cluster formation (and destruction) with other measurements related to the star formation efficiency such as the condensed baryon fraction (the fraction of the universally apportioned baryons within a dark matter halo that cool sufficiently below the halo virial temperature that they can be detected either as stars and cool gas, $f_{\rm C}$), the star formation efficiency as quantified by the ratio of stellar mass to cool gas mass, M$_{\rm G}$, or as the stellar mass to dark matter mass, M$_{\rm D}$. We have the necessary measurements of M$_{\rm G}$ and rotation speeds, v$_{\rm C}$, for a subset of our galaxies, mostly among the late types because we are utilizing the H{\small I} compilation of \cite{courtois11}. Both M$_{\rm G}$ and M$_{\rm D}$ are calculated as described in detail by \cite{zbtf}. We present these data in Figure \ref{fig:correlations} and the statistical significance of the resulting correlations between $\Delta {\rm T}_{\rm N}$ and these quantities (calculated by determining the probability that a random set of galaxies would have the same or larger Spearman rank correlation coefficient) in Table \ref{tab:spear}.

There is no correlation detected between $f_{\rm C}$ and $\Delta {\rm T}_{\rm N}$, and marginal correlations (at or slightly below 2$\sigma$ significance) between $\Delta {\rm T}_{\rm N}$ and the two measurements of star formation efficiency. With either measurement of star formation efficiency, the putative trend suggests that as the overall star formation efficiency increases (either relative to cold gas content or dark matter) the relative efficiency of cluster formation also increases. If a galaxy (of a given current-day stellar mass) has been more efficient at turning its baryons into stars it has also been more efficient at forming (or retaining) its clusters.

It is always difficult to interpret correlations, particularly among parameters that contain a measurement in common.  In this case the difficulty is that all of the quantities under discussion depend on M$_*$. To examine the sense of the effect expected in terms of parameter correlations should M$_*$ be incorrectly measured, consider that if M$_*$ is incorrectly overestimated (underestimated) then T$_{\rm N}$ will be underestimated (overestimated) and the star formation efficiencies will be overestimated (underestimated). The results of this behavior is that one would find cluster efficiency going in the opposite sense as star formation efficiency, opposite in sense to that observed. This argument suggests that our results are not due to correlated errors among the parameters arising from their common use of M$_*$. In fact, errors in M$_*$ may be weakening a stronger underlying correlation of the sense we observe.

\begin{figure}
\epsscale{0.9}
\plotone{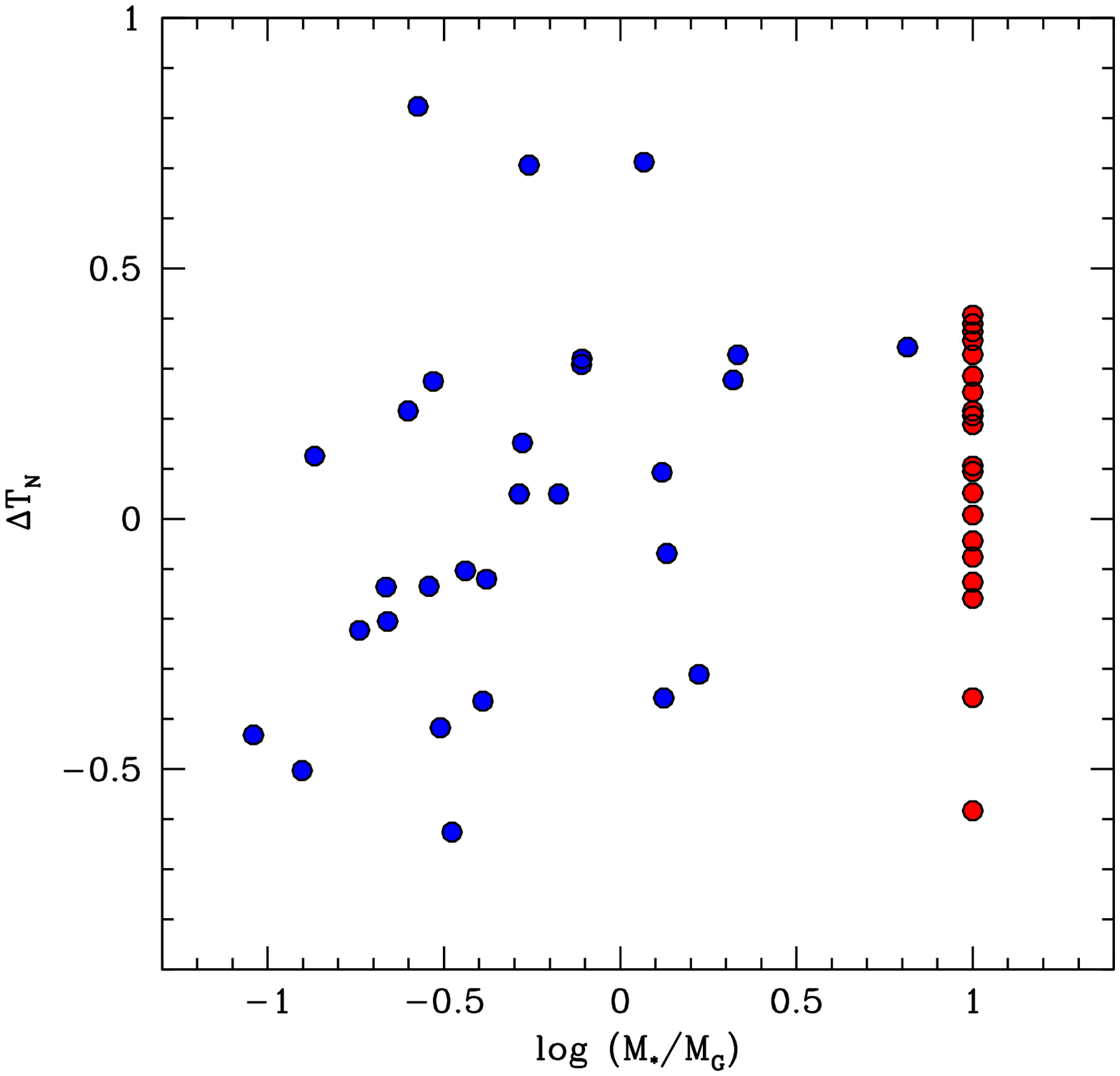}
\caption{The residuals about the mean trend for stellar mass normalized specific frequency, $\Delta {\rm T}_{\rm N}$, vs. the ratio of mass in stars to cold gas. The blue points represent the galaxies with $8.5 < \log ({\rm M}_*/{\rm M}_\odot) < 10.5$ in the sample for which we have H{\small I} measurements from the Cosmic Flows database \citep{courtois11}. The probability that a randomly chosen sample would have a Spearman rank correlation coefficient as large or larger than for the blue points is only 0.021. The red points represent early-type galaxies, T-type $< -2$, for which we do not have H{\small I} measurements, but for which we presume a small to negligible cold gas component, thereby placing these galaxies somewhere to the right of the blue points in this plot. Their position along the abscissa is arbitrarily set to $1$. As expected from the putative correlation, these stellar mass dominated systems tend to have $\Delta {\rm T}_{\rm N} > 0$, offering further evidence in favor of the existence of a correlation.}
\label{fig:correlations1}
\end{figure}

Even so, the statistical significance of the measured correlations is modest. To
test these correlations further we examine the behavior of our early type galaxies. For most of these galaxies we have no H{\small  I} measurement and hence also no measurement of v$_{\rm C}$, which is why these galaxies are absent from Figure \ref{fig:correlations}. If we presume that the gas content is a small fraction of the stellar content, we can place these galaxies at the right hand side of a revised version of the rightmost panel of Figure \ref{fig:correlations}, which we do in Figure \ref{fig:correlations1}.
Consistent with the expectation from the suggested correlation, most of these galaxies have $\Delta {\rm T}_{\rm N} > 0$. We conclude that the existence of the two correlations relative to different estimators of star formation efficiency plus the properties of the early-type galaxies all support the suggestion that T$_{\rm N}$ rises { relative to the mean} for galaxies that have converted a larger fraction of their baryons into stars.

Finally, we address the question of whether a more fundamental specific frequency is obtained when one normalizes relative to stellar mass or total mass. The latter normalization could be appropriate not necessarily because cluster formation somehow involves dark matter but because it depends on a quantity that scales more closely to dark matter than to stellar mass. One likely such candidate quantity is the baryonic mass of a galaxy. \cite{harris13} explored this question and found that using a dynamical estimator of halo mass did indeed lead to a tighter relationship between specific frequency and galaxy properties. With our sample we can only do this test with the limited subset of galaxies for which we have a measured v$_{\rm C}$. We find that our correlation is stronger when using M$_*$ (a Spearman rank correlation probability of arising randomly of 0.029) rather than with halo mass (comparable probability of 0.11). We conclude that we do not find support in our data for expressing specific frequency as a function of total mass, but note that we have a limited sample of only 29 galaxies currently with which we can do this test. This question merits further investigation.

\section{Summary}

The S$^4$G images \citep{sheth} are providing another opportunity to explore the bulk properties of globular cluster populations in galaxies. In Paper I we presented and validated our methodology and results for 97 early type galaxies. Here we present our results for a sample of  73 edge-on late type galaxies.

Using the combination of these two set of galaxies we find the following:

$\bullet$ We confirm previous findings \citep{georgiev,harris13} that the stellar mass normalized specific frequency decreases with stellar mass for intermediate mass galaxies (for our data that range is $8.5 < \log {\rm M}_*/{\rm M}_\odot < 10.5$) and flattens at higher stellar masses. The specific frequency may rise at even higher stellar masses, as measured by \cite{peng08}, \cite{harris13}, and ourselves in Paper I, but the new sample of galaxies presented here contains no galaxies in that mass range and therefore adds nothing to the existing discussion regarding the highest masses. Over the intermediate mass range, we find that T$_{\rm N} = 10^{6.7}{\rm M}_*^{-0.56}$. It is possible, given only our three galaxies with M$_* < 10^{8.5} {\rm M}_\odot$, that T$_{\rm N}$ is significantly larger for galaxies with lower stellar masses.

$\bullet$ We calculate that the cluster stellar mass fraction, the fraction of all stars in a galaxy that are in globular clusters, is 0.013 at the low end of the our sample's mass range and decreases to 10$^{-3}$ by a parent stellar mass of $10^{10.5} {\rm M}_\odot$. We speculate that this trend reflects the increased destruction rate of clusters in more massive systems { (see \cite{mieske} for a treatment of this topic that includes cluster population orbital properties as well as host mass}. Even so, clusters in the local Universe tend to be found around massive galaxies, as shown by our calculated globular cluster parent galaxy distribution function, G$({\rm M}_*)$. We present a fitting function for G$({\rm M}_*)$, G$({\rm M}_*) =  0.9 \exp (-({\rm M}_*/10^{11.3})^{2.3}) ((\log {\rm M}_* - 8.4)^{2.8} + 6.64)$ that can be used to calculate numbers of globular clusters hosted by galaxies of M$_* > 10^{8.5}$ M$_\odot$. 

$\bullet$ 
We find that the residuals of T$_{\rm N}$ about the mean trend do not correlate with a coarse measurement of large scale environment or morphology, but do appear to track the efficiency with which a galaxy has converted its baryons to stars. This efficiency can be quantified either as the ratio of stellar mass to dark matter mass or stellar mass to cold gas mass. In both cases, we find moderate correlations with { deviations in} T$_{\rm N}$. We are limited in that we do not have gas masses or rotation velocities for the entire sample. However, if we presume that the early type galaxies from Paper I have negligible cold gas masses, we find that these galaxies also follow the suggested correlation between { deviations} T$_{\rm N}$ { and} the ratio of stellar to cold gas mass. As such, we conclude that we find compelling evidence of a relationship that requires further investigation. Such a trend could inform models of different modes of global star formation in galaxies, where cluster formation, as well as other properties such as the stellar initial mass function, are impacted by the nature of star formation.

The bulk properties of globular cluster exhibit behavior that is not yet understood. As such, cluster populations provide another view into the complex history of galaxies and therefore an important constraint that should not be neglected when considering detailed models of galaxy formation and evolution.

\begin{acknowledgments}

DZ acknowledges financial support from 
NASA ADAP NNX12AE27G. KM acknowledges support through a NASA Spacegrant undergraduate research fellowship.
LCH acknowledges support from the Kavli Foundation, Peking University, and the Chinese Academy of Science through grant No. XDB09030102 (Emergence of Cosmological Structures) from the Strategic Priority Research Program.  
JHK acknowledges financial support from the Spanish Ministry of Economy and Competitiveness (MINECO) under grant number AYA2013-41243-P.
K.Sheth, J.C. Mu\~noz-Mateos, and T. Kim gratefully acknowledge support from the National Radio Astronomy Observatory, which is a facility of the National Science Foundation operated under cooperative agreement by Associated Universities, Inc. 
EA and AB acknowledge financial support from the CNES
(Centre National d?Etudes Spatiales - France). We also acknowledge support  from the People Programme (Marie Curie
Actions) of the European Union's Seventh Framework Programme FP7/2007-2013/ under
REA grant agreement number PITN-GA-2011-289313 to the DAGAL network.
This research has made use of the NASA/IPAC Extragalactic Database (NED), which is operated by the Jet Propulsion Laboratory, California Institute of Technology, under contract with NASA. 
\end{acknowledgments}

\begin{deluxetable*}{lrrrrrrrrrr}
\tablecaption{Globular Cluster Population Properties}
\tablewidth{0pt}
\tablehead{
\colhead{Name}&
\colhead{DM}&
\colhead{T}&
\colhead{m$_{3.6}$}&
\colhead{m$_{4.5}$}&
\colhead{N$_{50}$}&
\colhead{T$_{\rm N}$}&
\colhead{Background}&
\colhead{Q}\\
}
\startdata
ESO 107-016 & 32.26 & 8 & 15.3 & 15.7 &  129 &192$_ {-164}^{+320}$&$-$4.64&1 \\
ESO 146-014 & 31.64 & 8 & 15.2 & 15.8 &   27  &35$_{-32}^{+99}$&$-$5.26&1\\
ESO 249-035 & 31.77 & 8 & 17.0 & 17.5 &   75  &627$_{-607}^{+962}$&$-$4.92&0 \\
ESO 292-014 & 32.24 & 7 & 13.1 & 13.5 & 403  &68$_{-36}^{+49}$&$-$5.14&1\\
ESO 346-001 & 32.10 & 7 & 13.4 & 13.7 &   38  &11$_{-9}^{+49}$&$-$4.78&1 \\
ESO 356-018 & 31.59 & 9 & 15.0 & 15.5  &  81  &137$_{-67}^{+96}$&$-5.07$&1\\
\enddata
\label{tab:results}
\tablecomments{
DM refers to the distance modulus. T is the morphological T$-$Type of the galaxy from the compilation of \cite{buta15}. The 3.6 and 4.6$\mu$m magnitudes of the galaxies as measured by \cite{munoz} are in columns 4 and 5. N$_{50}$ is the number of globular clusters estimated from our best fit model of fixed power-law slope within 50 kpc of the galaxy. T$_{\rm N}$ is the number of these clusters per 10$^9$ M$_\odot$ of stellar mass in the galaxy. Background is the logarithm of the surface number density of background objects. The quality flag Q is defined by whether the data extend sufficiently over the radial range of interest to provide a robust constraint on the fitted models (Q$= 1$ is good, Q$= 0$ is poor).
This table is available in its entirety in a machine-readable form in the online journal. A portion is shown here for guidance regarding its form and content.}
\end{deluxetable*}

\begin{deluxetable*}{lr}
\tablecaption{Spearman Correlation Results : Probability of Being Randomly Drawn}
\tablewidth{0pt}
\tablehead{
\colhead{Parameters}&\colhead{P$_{R}$}
\\
}
\startdata
$\Delta {\rm T}_{\rm N}-f_{\rm C}$&0.692\\
$\Delta {\rm T}_{\rm N}$-$({\rm M}_*/{\rm M}_{\rm D})$&0.045\\
$\Delta {\rm T}_{\rm N}$-$({\rm M}_*/{\rm M}_{\rm G})$&0.021\\
T$_{\rm N}-{\rm M}_*$&0.029\\
T$_{\rm N}-{\rm M}_{\rm D}$&0.110\\

\enddata
\label{tab:spear}
\end{deluxetable*}

\end{document}